\documentstyle[amssymb, 12pt]{article}
\newcommand{\lab}[1]{\label{#1}}

\newcommand{\beq}{\begin{equation}}
\newcommand{\eeq}{\end{equation}}
\newcommand{\bea}{\begin{eqnarray}}
\newcommand{\eea}{\end{eqnarray}}
\newcommand{\sgn}{{\,\hbox{\rm sgn}\,}}
\newcommand{\tr}{{\,\hbox{\rm Tr}\,}}
\newcommand{\ad}{{\,\hbox{\rm ad}}}
\newcommand{\Ad}{{\,\hbox{\rm Ad}}}
\newcommand{\la}{\left\langle}
\newcommand{\ra}{\right\rangle}
\newcommand{\Hol}{{\,\hbox{\rm Hol}\,}}

\newcommand{\rf}[1]{{(\ref{#1})}}

\newcommand{\reali}{\mathinner{\bf R}}
\newcommand{\complessi}{\mathinner{\bf C}}
\newcommand{\toro}{\mathinner{\bf T}}
\newcommand{\interi}{\mathinner{\bf Z}}

\newcommand{\lieg}{{\frak{g}}}
\newcommand{\Ascr}{{\cal A}}

\newcommand{\Dscr}{{\cal D}}

\newcommand{\Gscr}{{\cal G}}
\newcommand{\Hscr}{{\cal H}}

\newcommand{\Oscr}{{\cal O}}
\newcommand{\Pscr}{{\cal P}}

\newcommand{\lora}{\longrightarrow}

\newcommand{\lolra}{\longleftrightarrow}

\newcommand{\I}{\mathop{\bf I}}
\newcommand{\sw}{\mathop{\bf P}}

\newcommand{\xnot}{{x_0}}

\begin{document}
\begin{titlepage}
\title{Topological $BF$ theories
in 3 and 4 dimensions\footnote{Work supported by
grants from MURST and by INFN}}
\author{
Alberto S. Cattaneo\\
Dipartimento di Fisica, Universit\`a di Milano\\
and I.N.F.N., Sezione di Milano\\
via Celoria 16, I-20133 {\sc Milano, Italy}\\
E-mail: {\em cattaneo@vaxmi.mi.infn.it}
\and
Paolo Cotta-Ramusino\\
Dipartimento di Matematica, Universit\`a di Milano\\
and I.N.F.N., Sezione di Milano\\
via Saldini 50, I-20133 {\sc Milano, Italy}\\
E-mail: {\em cotta@vaxmi.mi.infn.it}
\and
J\"urg Fr\"ohlich\\
Institut f\"ur Theoretische Physik, E.T.H.\\
H\"onggerberg, CH-8093 {\sc Z\"urich, Switzerland}\\
E-mail: {\em juerg@itp.ethz.ch}
\and
Maurizio Martellini\\
Dipartimento di Fisica, Universit\`a di Milano\\
and I.N.F.N., Sezione di Pavia\\
via Celoria 16, I-20133 {\sc Milano, Italy}\\
E-mail: {\em martellini@vaxmi.mi.infn.it}
}
\date{May 3, 1995 \\hep-th/9505027, IFUM 503/FT\\
prepared for the special
issue of {\em Journal
of Math.\ Phys.} on  {\em Quantum geometry and diffeomorphism-invariant
quantum field theory}\\
P.A.C.S. 02.40, 11.15, 04.60}

\maketitle

\begin{abstract}
\noindent
In this paper we discuss topological $BF$ theories in 3 and 4 dimensions.
Observables are associated to ordinary knots and links (in 3
dimensions) and to 2-knots (in 4 dimensions). The
vacuum
expectation values of such observables
give a wide range of invariants. Here we consider mainly
the 3 dimensional case, where
these invariants include Alexander polynomials, HOMFLY polynomials and
Kontsevich integrals.
\end{abstract}
\thispagestyle{empty}
\end{titlepage}
\newpage
\section*{I. Introduction}
This paper deals with a special kind of topological quantum field theories,
the $BF$ theories, that is the only known topological field theories
that can, in principle, be defined on a
a manifold $M$ of any dimension.
The symbol $BF$   means that the action contains
a term given by the wedge-product of an $(n-2)$-form $B$ of the adjoint type
times the curvature $F$ of a connection $A$. Here we have set $n=\dim M$.

Topological field theories
have been systematically considered by Witten
\cite{Wit1}, but somehow appeared in the literature
much before \cite{Schw}.

In a celebrated paper of 1989, Witten \cite{Wit2} showed that it is
possible to recover, via topological quantum field theories, the
invariants of links and knots known as Jones and HOMFLY polynomials
(\cite{Jones, HOMFLY, Lic, Weber}). The key idea was to consider a special
observable (``Wilson loop")
associated to knots and links and compute its vacuum expectation
value (v.e.v.) with respect to the Chern--Simons theory.

Let us now specifically consider $BF$ theories
(for an introduction to such theories see \cite{Bla}) and ask ourselves
what kind of topological invariants can we recover.

In an $n$-dimensional $BF$ theory, it is natural to look for observables
associated to imbedded (or immersed) manifolds in $M$ of codimension 2.
The v.e.v. with respect to $BF$ theory,
should then give topological invariants of these imbeddings (higher
dimensional knots).

Even though
the more interesting part of this program is related to the 4-dimensional
case, we have only systematically developed, up to now, the 3-dimensional
case, namely the case of ordinary knots and links. We have some hint
and some preliminary computations
about
the 4-dimensional case (see \cite{CM}), but most of
the work has yet to be done and, in this paper,
the 4-dimensional case is only briefly
sketched.

In the 3-dimensional case, we realized that, surprisingly enough,
$BF$ theory can significantly improve, in comparison with Chern--Simons
theory, our understanding
of the relation between
quantum field theory and knot invariants. Moreover, it
allows us to recover, as v.e.v.'s, some knot invariants that previously
have not been associated to quantum field theories (Alexander--Conway
polynomials)\cite{CCM}.

The original approach of Witten
to Chern--Simons field theory put the main emphasis on the non-perturbative
treatment. Here, instead, we stress the r\^ole of perturbation expansions
in the construction of knot invariants. The use of perturbative methods
is akin to the Vassil'ev approach to knot theory. More to
the point: the coefficients
of the perturbative series of topological field theories are precisely
knot invariants of finite type.

We have essentially two kinds of $BF$ theories. The first kind is what is
called $BF$ theory with a cosmological constant. The action is given by
the difference of two Chern--Simons actions (computed for two
different connections $A+\kappa B$ and $A-\kappa B$), where $A$ is a
connection,
$\kappa$ is a real parameter and $B$ is a 1-form of the adjoint type.
Pure
$BF$ theory is related to the Turaev-Viro \cite{TV} invariants just as
pure Chern--Simons theory is related to Reshetikhin-Turaev \cite{RT}
invariants.

In the $BF$ theory with a cosmological
constant, the observable to be associated to a knot is the trace of the
holonomy of the connection $A+\kappa B$ expanded in a Taylor
series in the variable $\kappa B$, at $\kappa=0$.

In $BF$ theory, the fields that are canonically conjugate are $A$ and $B$,
instead of $A$ being conjugate to itself (as in Chern--Simons). Hence
only contractions between the fields $A$ and $B$ are to
be considered. This feature
of the $BF$ theory is a very good one, since by considering, at the same
time, the Taylor expansion (in $\kappa B$) and
the vertex insertions (each
of them containing a factor multiplied by $\kappa^2$), one is able
to keep track of the various contributions
order by order in the variable $\kappa$.
This provides a much better control
of the perturbation series than in Chern--Simons theory.

The $BF$ theory with a cosmological constant leads to the same
knot invariants (HOMFLY and Jones polynomials) considered in Chern--Simons
theory.
But different choices of gauge produce different perturbation expansions,
i.e., different sequences of Vassil'ev invariants of knots, but associated
to the same knot polynomial. In this respect, let us point out that
suitable
normalization factors are to be taken into account, before one can
show that different perturbation expansions lead to the same knot polynomials.

The scheme for $BF$ theories is roughly as follows:
\begin{itemize}
\item {\it Covariant gauge.} In this
gauge the terms of the perturbative expansion
are multiple linking-integrals.
\item {\it Holomorphic gauge.} In this gauge the terms of the perturbative
expansion are Kontsevich integrals.
\item {\it Axial gauge.} In this gauge the terms of the perturbative expansion
are expressed as  sums of ``tensors" over
the set of vertices of a given projection of the
knot.
\end{itemize}

The second kind of 3-dimensional
$BF$ theory that we consider is the one {\it without} cosmological constant.
The action here is given by the {\it derivative}
of the Chern--Simons action computed for the connection
$A+ \kappa B$ at $\kappa=0$. The observable
is an exponential function of the derivative of the holonomy.

The perturbative expansion of the
$BF$ theory without cosmological constant produces the
coefficients of the Alexander--Conway
polynomial. They  cannot be recovered in the framework of Chern--Simons
field theory.

Finally, concerning the higher dimensional $BF$ theories,
it is very likely that invariants of 2-knots as
well as invariants of 4-manifolds can be recovered in the framework
of such theory. In this respect $BF$ theories can play
a r\^ole in the loop-variables formulation of quantum gravity.

\section*{II. Geometry of $BF$ theories}
Topological $BF$ theories are the only known
topological quantum field theories that can be consistently defined
in any dimension. Thus we
consider a (compact, oriented, closed, Riemannian)
manifold $M$ of dimension $n$,  with a $G$-principal bundle
$P\lora M$.
Here $G$ is a compact simple Lie group with Lie algebra $\lieg$. We will mainly
consider $G=SU(N)$.

Let us denote by $\Omega^*(M)$
the space of differential forms on $M$ and by
$\Omega^*(M,\ad P)$ the space of differential
forms on
$M$ with values in the adjoint bundle $\ad P\equiv P\times_{\Ad} \lieg$
(locally $\lieg$-valued forms on $M$).
\\
On $M$ we can consider a quantum field theory depending on
two fields:
\begin{itemize}
\item the connection $A$ (with curvature denoted by
$F_A$, or simply by $F$, that is a form
in $\Omega^2(M,\ad P)$)
\item a form $B\in \Omega^{n-2}(M,\ad P)$
\end{itemize}
With the above ingredients we can construct an action
\beq
S_{BF} = \int_{M} \tr(B\wedge F),
\lab{bf}
\eeq
where the trace refers to an assigned representation of $\lieg$.
Most commonly,
we will consider the fundamental
representation.
The corresponding Gibbs measure will be given by
$\exp (ifS_{BF})$ where $f$ denotes a coupling constant.

We denote by $\Gscr$ the group
of gauge transformations. For any $\psi\in \Gscr$,
locally given by a map $\psi:M
\mapsto G$, the field $B$ transforms
as $B\lora \psi^{-1}B\psi$. The action \rf{bf} is then
gauge invariant. Moreover it is invariant under diffeomorphisms
(being given by the integral of a $n$-form) and it is independent
of the metric in $M$. In other words, the action \rf{bf} defines,
in principle,
a {\it topological field theory in any dimension}.

In 3 and 4 dimensions we can study other types of $BF$ action.
Namely for any values of the
parameter $\kappa$ we can consider, in 3 dimensions,
the action:
\beq
S_{BF,\kappa} =  \int_{M} \tr(B\wedge F) +{{\kappa^2}\over 3}\int_M
\tr (B\wedge B\wedge B),
\lab{bf3}
\eeq
and, in 4 dimensions, the action:
\beq
S_{BF,\kappa} = \int_M \tr(B\wedge F) +{{\kappa}\over 2} \int_M
\tr (B\wedge B).
\lab{bf4}
\eeq
In order to understand the geometrical significance of the above
actions, we recall that in 4 dimensions there is a
topological invariant represented by the integral of
the Chern--Weil form:
\beq
Q_2(F)\equiv\int_M \tr(F\wedge F)
\lab{Q2}
\eeq
while, in 3 dimensions, we have the secondary topological
invariant, locally represented
by the integral of the Chern--Simons form:
\beq
S_{CS}(A)\equiv \int_M \tr (A\wedge dA +{2\over3}A\wedge A\wedge A).
\lab{CS}
\eeq
The actions \rf{bf} (with $\dim M=3,4$), \rf{bf3} and \rf{bf4}
are all {\it variants} of the above topological invariants.

More precisely
the following simple relations hold in dimensions 3 and 4 respectively:
\beq
\begin{array}{ll}
(1/2)[S_{CS}(A+\kappa B)-S_{CS}(A-\kappa B)] &= 2\kappa S_{BF,\kappa}\\
\displaystyle{{d\over{d\kappa}}
S_{CS}(A+\kappa B)\bigg|_{\kappa=0}} &= 2 S_{BF} \quad (\dim M=3)
\end{array}
\lab{var3}
\eeq
and
\beq
\begin{array}{ll}
Q_2(F+\kappa B)-Q_2(F)&= 2\kappa S_{BF,\kappa},\\
\displaystyle{{d\over{d\kappa}}
Q_2(F+\kappa B)\bigg|_{\kappa=0}}&=2S_{BF}\quad (\dim M=4)
\end{array}
\lab{var4}
\eeq
The action \rf{bf3} (and sometimes also the action \rf{bf4}) is called
$BF$ action {\it with a cosmological term}. The reason for this terminology
is easily explained: let us consider in 3 dimensions,
the frame bundle $LM$ (with group
$G=GL(3)$).
The soldering form $\theta$ is a 1-form with values
in $\reali^3$ associated to the fundamental representation of
$GL(3)$. When  $\theta$ is restricted to the orthonormal frame bundle
and is expressed in local coordinates we obtain the ``dreibein"
$\{e^i\}_{i=1,2,3}.$

In the so-called first-order formalism, the classical action for
gravity is given by
\beq
\int_M \sum_{i,j,k}\epsilon_{ijk}e^i \wedge R^{jk} +\kappa^2 \int_M
\sum_{i,j,k}
\epsilon_{ijk}
e^i\wedge e^j\wedge e^k
\lab{grav}
\eeq
where the matrix $R$ is the curvature 2-form and the second term in the
integral is the cosmological term ($\kappa^2$ is
the cosmological constant). In 3 dimensions we can consider
the linear
isomorphism $\reali^3\mapsto Lie SO(3)$ given by $e_i\lora \sum_{j,k}
\epsilon_{ijk}
(E^j_k-E^k_j)$ where $e_i$ are the elements of a canonical basis of $\reali^3$
and $E^j_k$ is the matrix whose $(m,n)$-entry is
given by $\delta^{j,m}\delta_{k,n}$.
Under this isomorphism, the soldering from is transformed into the
$B$-field, and \rf{grav}
becomes, up to a constant, the $BF$ action \rf{bf3}.
\\
Next we discuss the {\it symmetries} of the $BF$ theories.
First of all we have to consider the group of
gauge transformations, whose infinitesimal
action on the fields is given by:
\beq
A\lora A+d_A\xi;
\quad B\lora B+ [B,\xi].
\lab{gaugetrans}
\eeq
Here the infinitesimal gauge transformation $\xi$
is an element of $\Omega^0(M,\ad P)$.

In $BF$ theories there exists another important set of symmetries.
In this regard, we have to distinguish between
the 3-dimensional and the 4-dimensional
case.

In 3 dimensions the action \rf{bf3} is invariant
also under the following infinitesimal transformations:
\beq
A\lora A + \kappa^2[B,\chi],
\quad B\lora B+ d_A\chi,
\lab{bftrans3}
\eeq
where $\chi\in\Omega^0(M,\ad P)$ is an infinitesimal gauge
transformation
(in general different from $\xi$).
Instead  in 4 dimensions \rf{bf4} is invariant under the
following infinitesimal transformations:
\beq
A\lora A+ \kappa\eta,
\quad B\lora
B-d_A \eta,
\lab{bftrans4}
\eeq
where  $\eta$ is a form in $\Omega^1(M,\ad P)$,
i.e. is the difference of
two connections.

The geometrical meaning of the combination of
transformations \rf{gaugetrans} and \rf{bftrans3} is
straightforward; when $\kappa\neq 0$,
\rf{gaugetrans} and \rf{bftrans3} are {\it equivalent}
to the following infinitesimal gauge transformations:
\beq
\begin{array}{ll}
A+\kappa B\lora & A+\kappa B + d_{A+\kappa B} (\xi+\kappa\chi)\\
A-\kappa B\lora & A-\kappa B + d_{A-\kappa B} (\xi-\kappa\chi)\\
\lab{equiv}
\end{array}
\eeq
when $\kappa=0$, \rf{gaugetrans} and \rf{bftrans3} are equivalent to
the two sets of transformations obtained by
\begin{enumerate}
\item
evaluating both sides of \rf{equiv} at
$\kappa=0$ and
\item applying the operator
$\displaystyle{{d\over{d\kappa}}\bigg|_{\kappa=0}}$ to both sides of
\rf{equiv}.
\end{enumerate}
It is important to remark that in $BF$ theory (with $\kappa=0$
and with $\kappa\neq 0$) we have two distinct
infinitesimal gauge transformations
$\xi$ and $\chi$ that generate the symmetries. In the corresponding
quantum theory, this  implies that there are two distinct set of ghosts
that will produce cancellations in the perturbative expansion.

In 4 dimensions
the invariance of \rf{bf4} (with $\kappa \neq 0$)
under \rf{bftrans4} is
nothing else but the independence of the $BF$ action
of the
connection
$A$. In this way we ensure that
the 4-dimensional
$BF$ action has the same kind of symmetries
as the Chern--Weil form.

When $\kappa=0$ the 4-dimensional $BF$ action is,
in general, not independent
of $A$ anymore and the invariance under the
transformation $B\lora B+ d_A\eta$ is simply
a consequence of the Bianchi identity.

In contrast to the
3-dimensional case,  the two sets of ghosts
generated by
the invariance under \rf{gaugetrans} and \rf{bftrans4} have a different
nature (0-forms vs. 1-forms).

\section*{III. $BF$ observables in 3 dimensions}
The fundamental fields of our theory (in an $n$-dimensional
manifold) are the connection 1-form $A$
and the $(n-2)$-form $B$.
This suggests that the right observables for
the topological $BF$ theories should be associated to collections
of loops (links) in $M$ (i.e.\ one-dimensional submanifolds)
and to $(n-2)$-submanifolds.

Before discussing the precise definition of our observables, let us consider
the case of an abelian $BF$ theory in $S^n$ that is both simple and
instructive.

The action for such a theory is given by
$S_{BF}=\displaystyle{\int_{S^n} B\wedge dA}$. This action is invariant under
the transformations: $A\lora A+d\xi, \quad B\lora B+ d\eta,$  where
$\xi \in\Omega^0(M)$ and $\eta \in \Omega^{n-2}(M)$.

In such a theory we can associate an observable to any imbedded oriented loop
$C$
and an observable to any imbedded oriented closed $n-2$ submanifold $S$.

These observables are given by $\Oscr_1(C)\equiv \displaystyle{\int_C A}$ and
$\Oscr_{n-2}(S)\equiv \displaystyle{\int_S B}$. They are obviously
invariant under the symmetries of $BF$ theory.

The holonomy along an embedded circle $C$
is given, in the abelian case, by
\[
\displaystyle{
\Hol(A;C)= \exp \left(\Oscr_1(C)\right)
}.
\]
Since only the kinetic term $B\wedge dA$
appears in the
lagrangian, we only have to consider
{\it vacuum expectation values} (v.e.v.) of the form:
\beq
\la A_{\nu}(x)B_{\mu_1,\mu_2,\cdots,\mu_{n-2}}(y)\ra=
{{1}\over {if \Omega_n}}
\sum_k\epsilon_{\nu,{\mu_1},{\mu_2},\cdots,{\mu_{n-2}},k}
\displaystyle {{x_k-y_k}\over
{||x-y||^n}};
\lab{AB}
\eeq
where $\Omega_n$ is the ``volume" of the unit sphere in $n$ dimensions, and
$x_k$
are the coordinates of $x$,
Hence only observables with a number of $A$-fields equal to
the number of $B$-fields will have non-vanishing expectation values.

In particular:
\[
if <\Oscr_1(C)\Oscr_{n-2}(S)>=\int_C\int_S dx_\nu dy_{\mu_1}dy_{\mu_2}\cdots
dy_{\mu_{n-2}}\la A_{\nu}(x)B_{{\mu_1},\cdots,{\mu_{n-2}}}(y)\ra=lk(C,S)
\]
where $lk(C,S)$ is the (higher-dimensional) linking number between $C$ and $S$.

The v.e.v. of all the observables one can consider in the abelian theory
are thus given by functions of linking numbers between loops and
closed $(n-2)$-submanifolds.

The non-abelian theory is more complicated. To each loop
$C$ we can still associate the relevant (trace of the) holonomy of the
connection $A$. As we will discuss below,
to each $(n-2)$-dimensional imbedded submanifold $S$ we can associate
an observable closely related to the integral over $S$ of the
$(n-2)$-form $B$.

Since the kinetic part of the non-abelian theory is the same
as the one of the abelian one, \rf{AB} still holds in the
slightly modified form:
\beq
\la A^a_{\nu}(x)B^b_{\mu_1,{\mu_2},\cdots,{\mu_{n-2}}}(y)\ra=
\delta_{a,b} {{1}\over {if \Omega_n}} \sum_k
\epsilon_{\nu,{\mu_1},{\mu_2},\cdots,{\mu_{n-2}},k}\displaystyle
{{x_k-y_k}\over
{||x-y||^n}}.
\lab{nonabAB}
\eeq

Moreover, in a non abelian theory, vertex terms are present; so
we have other non trivial v.e.v.'s like
\[
\la A^a(x) B^b(y) B^c(z)\ra;\quad \la A^a(x) A^b(y) A^c(z)\ra
\]
that will produce multiple integrals of (convolutions) of
the same kernels that appear in \rf{nonabAB} (iterated linking numbers).
Here neither we have v.e.v.'s of the type:
\[
\la A^a(x) A^b(y) B^c(z)\ra
\]
nor we have ``loops" since loops
are cancelled by the corresponding diagrams involving
ghosts.

In this way, non abelian gauge theories yield invariants
associated to $(n-2)$-submanifolds and imbedded circles that are more
sophisticated
than the invariants related to the abelian theory.

In this paper we
are mainly interested in the case $n=3$, so imbedded $(n-2)$-submanifolds
are knots. The 3-dimensional $BF$ theory then becomes a theory of
links in a 3-dimensional manifold.

We now consider the precise definition of our observables.
In the framework of
{\it 3-dimensional $BF$ theory with a cosmological constant},
the natural observables to be associated to a knot $C$ are given
by
\beq
\tr\Hol(A \pm \kappa B;C)
\lab{obsbfcc}
\eeq
while the natural observable to be associated to a knot $C$
in a {\it 3-dimensional
$BF$theory without cosmological constant} is given by
\beq
\displaystyle{{d\over{d\kappa}}\bigg|_{\kappa=0} \tr \Hol(A+\kappa B;C)
=\tr \int_C \Hol_{x_0}^y(A;C) B(y) \Hol_y^{x_0}(A;C) }.
\lab{obsbf}
\eeq
In this expression, $x_0\in C$ is a fixed point on the knot,
$\Hol_{x_0}^y(A;C)\equiv \Pscr \exp \int_{x_0}^y A$, where $\Pscr$
denotes path-ordering and  the integral is meant
to be computed along the arc of $C$ joining $x_0$ to $y$ in the direction
prescribed by the orientation of the knot. Given a section
$\sigma:M\lora P$, the group element $\Hol_{x_0}^y(A;C)$ can be equivalently
described by the equation $\sigma(y)\Hol_{x_0}^y(A,C)=C^h(y)$ where
$C^h$ denotes the horizontal lift of $C$ with starting point $\sigma{\xnot}$.
Also, by the symbol $\Hol_{\xnot}(A;C)$ we denote the holonomy
along $C$ with base point $\xnot.$\\

We now consider the Taylor expansion
of \rf{obsbfcc} at $\kappa=0$.
For this purpose we compute
$\gamma_n(C,\xnot)\equiv\displaystyle{{1\over{n!}}{{d^n}\over{d\kappa^n}}\bigg|_{\kappa=0}}
\Hol_{\xnot}(A+\kappa B;C),$
obtaining
\beq
\begin{array}{ll}
\gamma_0(C,\xnot)=\Hol_{\xnot}(A;C) \\
\gamma_1(C,\xnot)=\displaystyle{\int_C \Hol_{\xnot}^y B(y) \Hol_y^{\xnot}}\\
\gamma_2(C,\xnot)=\displaystyle{\int_{y_1<y_2\in C}\Hol_{\xnot}^{y_1} B(y_1)
\Hol_{y_1}^{y_2} B(y_2)\Hol_{y_2}^{\xnot}}\\
\cdots\cdots\\
\gamma_n(C,\xnot)=\displaystyle{\int_{y_1<\cdots<y_n\in C}\Hol_{\xnot}^{y_1}
B(y_1)
\Hol_{y_1}^{y_2} B(y_2)\cdots \Hol_{y_{n-1}}^{y_n} B(y_n)\Hol_{y_n}^{\xnot}}.\\
\end{array}
\lab{gammas}
\eeq
In our notation we do not write
explicitly the
dependence of $\gamma_n(C,\xnot)$ on $A$ and $B$.

The above formulas are {\it iterated Chen integrals}.
In fact, let us define
\beq
\hat B(x)\equiv \Hol_{\xnot}^x B(x) [\Hol_{\xnot}^x]^{-1}.
\lab{bhat}
\eeq
This is a $\lieg$-valued 1-form on $C$.
The geometrical meaning of $\hat B$ is as follows:
we can view the 1-form $B$ equivalently as an element of $\Omega^1(M,\ad P)$
or as a $\lieg$-valued 1-form on the total space of the principal
bundle $P(M,G)$ which is tensorial under the adjoint action. Given a reference
section $\sigma:M\lora P$, we can consider the horizontal lift $C^h$ of
$C$ with starting point $\sigma(\xnot)$. The integral of $B$ (seen as
a 1-form on $P$) along $C^h$ is exactly the integral of $\hat B$ along the loop
$C$.

The definitions \rf{gammas} coincide with the following Chen integrals
\cite{Chen}:
\beq
\gamma_n(C,\xnot)=\oint_{\xnot}
\underbrace{
\hat B\cdot\hat B\cdots\hat B \cdot \hat B}_{\mbox{$n-1$ times}}\cdot
(\hat B\Hol_{\xnot}(A;C))
\lab{chen}
\eeq
We recall that the iterated integral $\displaystyle{
\int_a^b \omega_1\cdot\omega_2
\cdots \omega_n}$ of $n$ 1-forms $\{\omega_i\}_{i=1,\cdots n}$ (with values
in any algebra) is given
(in our notation) by the formula
$\displaystyle{\int_{a<x_1<\cdots<x_n<b}
\omega_1(x_1)\wedge \omega_2(x_2)\wedge\cdots\wedge\omega_n(x_n)}.$
\\

Our Taylor expansion finally reads
\beq
\Hol_{\xnot}(A +\kappa B; C) = \sum_n \kappa^n \gamma_n(C,\xnot)
\lab{taylor}
\eeq
We may also try to consider
as observables the quantities
$\tr \gamma_n(C,\xnot)$. They  are all gauge
invariant, i.e. invariant under \rf{gaugetrans}, but unfortunately they are
not invariant under \rf{bftrans3}. In fact, under the transformations
\rf{bftrans3}
we have the
following transformation:
\beq
\gamma_n(C,\xnot)\lora\gamma_n(C,\xnot) +\kappa^2 \tilde\gamma_{n+1}(C,\xnot)
-\tilde \gamma_{n-1}(C,\xnot)-[\chi(\xnot),\gamma_{n-1}]
\lab{tilde}
\eeq
Here the map $\gamma\lora \tilde\gamma$ is meant to be the derivation that
replaces in \rf{gammas},  the field
$B$ evaluated at a given set of points $\{y_i\}$
by the field $[B,\chi]$ evaluated at the same points $y_i$.

As a consequence of the above transformation laws, we conclude that
only particular combinations of $\gamma_n(C,\xnot)$ give rise to
good observables.

Namely the observables that we can consider
for the $BF$ theory
{\it with} a cosmological constant are only
{\em the traces}
of the following quantities
\beq
\begin{array}{ll}
\Hol(A\pm \kappa B) \\
\Hol_{\xnot}^{even}(C)\equiv
(1/2)[\Hol_{\xnot}(A+\kappa B;C)+\Hol_{\xnot}(A-\kappa B;C)]=\sum_s\kappa^{2s}
\gamma_{2s}(C,\xnot)\\
\\
\Hol_{\xnot}^{odd}(C)\equiv
(1/2)[\Hol_{\xnot}(A+\kappa B;C)-\Hol_{\xnot}(A-\kappa
B;C)]=\sum_s\kappa^{2s+1}
\gamma_{2s+1}(C,\xnot).\\
\end{array}
\lab{taylor1}
\eeq
Moreover, as expected,
the $BF$ theory {\it without} cosmological constant (i.e., with $\kappa=0$),
admits as observables, either
$\tr[\gamma_i(C;\xnot)]$,
$i=0,1$  or
traces of products of $\gamma_i(C;\xnot),\;i=0,1.$

When we consider the last case, we have to allow only
infinitesimal transformations \rf{bftrans3} that satisfy the extra-condition
$\chi(\xnot)=0.$ In other words $\chi$ must belong to the Lie algebra
of the group of gauge transformations, whose restriction
to $\xnot$ is the identity.

In particular, we will be interested in the following set of observables
for the $BF$ theory {\it without} cosmological constant:
\beq
\Gamma_n(C,\xnot)={1\over{n!}}
\underbrace{
\oint_{\xnot}
\hat B\oint_{\xnot}\hat B\cdots
\oint_{\xnot}\hat B}_{\mbox{$n-1$ times}}
\oint_{\xnot}\hat B\Hol_{\xnot}(A;C)
\lab{chenab}
\eeq
The observables $\gamma_n$ and $\Gamma_n$ do not coincide,
since
$\hat B(x)$ and $\hat B(y)$ are {\em not} commuting quantities.
We now define
\beq
\Hscr(C;\lambda)\equiv \sum_n \lambda^n \Gamma_n(C;\xnot).
\lab{taylorab}
\eeq
The quantity
$\tr \Hscr(C;\lambda)$ replaces $\tr \Hol(A+\kappa B,C)$
as the {\it basic observable} for the $BF$ theory with {\it zero cosmological
constant}. The geometrical meaning of \rf{taylorab} is related to the action
of the group $G$ (or of its tangent bundle) on the tangent bundle
of the total space $P$ and will be discussed elsewhere.

\section*{IV. Formal relations between Chern--Simons and $BF$ theories}

Let us consider the Chern--Simons partition functions:
\beq
\begin{array}{ll}
Z_{CS}(M,C;k)&\equiv\displaystyle{
\int\Dscr A \exp\bigl(i k S_{CS}(A)\bigr)\tr\Hol(A;C)}\\
Z_{CS}(M;k)&\equiv
\displaystyle{\int \Dscr A \exp\bigl(i k S_{CS}(A)\bigr)}\\
\end{array}
\lab{CSac}
\eeq
and the $BF$ partition functions:
\beq
\begin{array}{ll}
Z_{BF,\kappa}(M,C;f)&\equiv\displaystyle{\int\Dscr A\Dscr B
\exp\bigr(i f S_{BF,\kappa}(A,B)\bigr)}
\tr\Hol(A+\kappa B)\\
Z_{BF,\kappa}(M;f)&\equiv \displaystyle{\int \Dscr A\Dscr B \exp\bigl(i f
S_{BF,\kappa}(A,B)
\bigr)}\\
\end{array}
\lab{BF}
\eeq
where $k$ and $f$ are coupling constants. The constant $k$ is quantized, namely
must be an integer multiple of ${(4\pi)^{-1}}$ in order to guarantee
the invariance of the action $S_{CS}$
under gauge transformations not connected to the identity.
At the formal level we have:
\beq
\displaystyle{
Z_{CS}(M;k)\overline{Z_{CS}(M;k)}
}=
Z_{BF,\kappa}(M;f).
\lab{cscs}
\eeq
In fact, the first term in the above equation is given by:
\[
\int \Dscr A_1\Dscr A_2 \exp\left[i k S_{CS}(A_1)-i k S_{CS}(A_2)\right],
\]
and this quantity is equal to $Z_{BF,\kappa}(M;f),$ provided we set:
\beq
2 A=A_1 +A_2;\quad 2\kappa B=A_1-A_2 ; \quad f= 4 \kappa k.
\lab{csbf}
\eeq
{\it Assuming that $Z_{CS}(M;k)$ represents the Reshetikhin--Turaev
\cite{RT}
invariant, then $Z_{BF,\kappa}(M,f)$ represents the Turaev--Viro
\cite{TV}
invariant}.
\\

Next we discuss the relations between the $BF$ and the CS actions with
{\it knots incorporated}. We require again relations
\rf{csbf}. Then we have that
\beq
Z_{CS}(M,C;k)\overline{Z_{CS}(M;k)}
=Z_{BF,\kappa}(M,C;f),
\lab{csbfk}
\eeq
and hence
\beq
\displaystyle{
{{Z_{CS}(M,C;k)}\over{Z_{CS}(M;k)}}}
=\displaystyle{
{{Z_{BF,\kappa}(M,C;f)}\over{Z_{BF,\kappa}(M;f)}}
}.
\lab{homfly}
\eeq
When we choose $M=S^3$ and  consider the fundamental representation
of $SU(N)$, then the normalized partition function
\rf{homfly} {\it gives (a regular isotopy
invariant version of) the HOMFLY polynomial $P(l,m)$}
evaluated at $l= \exp (-i f^{-1} \kappa N),\quad
m=l^{1/N}-l^{-1/N}.$ From now on we set
$f=(2\pi)^{-1}.$

The polynomial $P(l,m)$ satisfies the skein relation:
$lP(l,m)(C_+)-l^{-1}P(l,m)(C_-)=mP(l,m)C_0$, where $\{C_+,C_-,C_0\}$
is a Conway triple, and the normalization
condition $P(l,m)(\emptyset)=1$ for the empty knot $\emptyset$
is imposed.

\section*{V. Choice of gauge and v.e.v's}

In order to quantize $BF$ theory we first need to make a choice of
gauge. The most natural choice of gauge is the {\it (background) covariant
gauge}.
Namely, we fix a background connection $A_0$ and we require that the
fields $A$ and $B$ satisfy the following constraints:
\beq
d^*_{A_0}(A-A_0) = d^*_{A_0} B=0,
\lab{covgauge}
\eeq
where $d_{A_0}^*$ is the adjoint of the covariant derivative.
This is a complete gauge condition,
namely it provides us with a honest (local)
section of the
the bundle of gauge orbits $\displaystyle{\Ascr\lora \Ascr/\Gscr}$, where
$\Ascr$ denotes the space of all (irreducible) connections.

With this choice of gauge, we conclude that, for any equivalence class
of connections $[A]$, $[B]$ represent
a tangent vector in $\displaystyle{T_{A}\left(\Ascr/\Gscr\right)}$
the space of gauge orbits (or a cotangent vector if
use the Hodge star operator to introduce an inner product in
$\displaystyle{T_{A}\left(\Ascr/\Gscr\right)}.$

In physics one would like to choose the canonical
flat connection, as a background
connection, and hence replace the covariant derivative with
the exterior derivative. This is always possible in 3-dimensions,
when the group $G$ is $SU(N)$.
In this case, the {\em covariant gauge condition} read,
in local coordinates,
\[
\sum_{\mu}\partial^{\mu} A_{\mu}= \sum_{\mu}\partial^{\mu} B_{\mu}=0.
\]

When the 3-dimensional manifold is $\reali^3,$ or more generally
$\Sigma \times \reali,$ for a given surface $\Sigma$,
we can consider other gauges. These are
not true complete gauge conditions,
in the sense specified above, since, after
imposing them, we are left
with a residual
freedom in the choice of gauge.

For $M=\Sigma \times \reali$, we denote by $t$ the coordinate
of $\reali$.
We introduce a complex structure in $\Sigma$ (with local coordinates
$z=x_1+ix_2$, $\bar z=x_1-ix_2$). This yields a decomposition
of $\Omega^1(\Sigma,\ad P)$ into a holomorphic part $\Omega^{1,0}
(\Sigma,\ad P)$ and an anti-holomorphic part $\Omega^{0,1}
(\Sigma,\ad P)$ \cite{AB}.
By saying that we choose
the {\em light-cone gauge} in the {holomorphic} formulation, we mean
that we assume that, for each $t\in \reali$, both the connection
$A(t)$ and the 1-form $B(t),$ restricted to $\Sigma$, are {\it holomorphic}.
In other words, in local coordinates,
$A$ and $B$ are expressed as:
\[
A_zdz+A_0dt,\quad B_zdz + B_0dt.
\]
This choice of gauge is equivalent to requiring that,
 in {\em real} coordinates
$x_1,x_2,t$, we have $A_1=A_2$ and $B_1=B_2$.
In this  gauge the $BF$ action becomes:
\beq
S_{BF,\kappa}= \int_{M} \tr\left(B_z\wedge \bar \partial A_0 -B_0\wedge
\bar \partial A_z\right),
\lab{acthol}
\eeq
namely it is quadratic and independent of $\kappa$. The quantization of
Chern--Simons
theory in the light-cone gauge has been studied in
\cite{FK}.

Finally we consider the {\em axial gauge} $A_0=B_0=0$.
Here, again, the $BF$ action is quadratic and independent of $\kappa$:
\beq
S_{BF,\kappa}=\int_{M} \tr \left(B_2 \wedge d_0 A_1 - B_1 \wedge d_0
A_2\right).
\lab{actax}
\eeq
Let us consider the vacuum expectation values of the $BF$ theory in the
three different gauges defined above. In the two singular gauges, we only have
a quadratic kinetic term in the lagrangian; this implies that the
two-point correlation functions determine all $n$-point
correlations (by Wick's theorem). The v.e.v.'s in the
different gauges are as follows
\begin{itemize}
\item Covariant gauge ($M={\reali}^3$):
\beq
\begin{array}{ll}
\la A_{\mu}^a(x) B_{\nu}^b(y)\ra &=-4\pi i \delta^{ab}l(x,y)\\
\la A_{\mu}^a(x) B_{\nu}^b(y) B_{\rho}^c(z)\ra &=-(4\pi)^2
f^{abc}v_{\mu\nu\rho}
(x,y,z)\\
\la A^a_{\mu}(x) A_{\nu}^b(x) A_{\rho}^c(x)\ra &=-(4\pi)^2 \kappa^2 f^{abc}
v_{\mu\nu\rho}(x,y,z)\\
\end{array}
\lab{covev}
\eeq
where we have set:
\beq
\begin{array}{ll}
l_{\mu\nu}(x,y)&
\equiv\displaystyle{{1\over{4\pi}}\sum_{\rho}\epsilon_{\mu\nu\rho}\displaystyle
{
{
{x_{\rho}-
y_{\rho}}
\over
{||x-y||^3}
}
}}\\
\\
v_{\mu\nu\rho}(x,y,z)&\equiv\displaystyle{\int_{\reali^3}d^3w
\sum_{\alpha,\beta\gamma}\epsilon^{\alpha\beta\gamma}
l_{\mu\alpha}
(x,w) l_{\nu\beta}(y,w) l_{\rho\gamma}(z,w)}\\
\end{array}
\lab{lv}
\eeq
\item Holomorphic gauge ($M=\complessi \times \reali$):
\beq
\la A_z^a(z,t)B_0^b(w,s)\ra= \displaystyle{-2\delta^{ab}{1\over
{(z-w)}}}\delta(t-s)
\lab{hovev}
\eeq
\item Axial gauge ($M=\reali^3$):
\beq
\la A_1^a(x_1,x_2,x_0) B_2^b(y_1,y_2,y_0)\ra= -(2\pi i)\delta^{ab}
\sgn(x_0-y_0)\delta(x_1-y_1)
\delta(x_2-y_2).
\lab{axvev}
\eeq
\end{itemize}
\section*{VI. The framing of the knot and the skein relation}
We require that v.e.v.'s
involving the fields $A$ and
$B$ are not computed at coincident points. This is equivalent to requiring
that, in all the integrals of
the perturbative expansion
of v.e.v.'s of observables, the field $A$ lives on a companion knot $C_f$
of the original knot
$C$ where the field $B$ is supposed to be integrated over.
Thus we must consider a {\it framing} of the original knot.

We denote by $\epsilon$
the distance of the companion knot $C_f$ from $C$.
Eventually we will have to consider the limit $\epsilon
\to 0$, in order to restore the diffeomorphism-invariance broken by the
introduction of the framing.

Now we want to study the effect of a small deformation of the knot,
concentrated
around a given point $x$
of $C$ and, simultaneously, of its
companion $C_f$.

These deformations will change the holonomies by a factor
proportional to the curvature, i.e. they will modify the v.e.v. as follows:
\[
\la\tr\Hol(A+ \kappa B)(C)\ra\lora \la \tr\{
\Hol_{\xnot}^x (A+\kappa B)F_{A+\kappa B}(x)\Hol_x^{\xnot}(A+\kappa B)\}\ra.
\]
Notice that $F_{A+\kappa B}= F_A +\kappa d_AB + \kappa^2 B\wedge B$.
As in \cite{CGMM}, we {\it assume} that
we can perform an integration by parts.
In order to do so, we
first compute the functional derivatives of the $BF$ action, obtaining:
\beq
\begin{array}{ll}
\displaystyle{{{\delta S_{BF,\kappa}}\over{\delta A^a_{\mu}(x)}}}=
\displaystyle{{1\over{4}}\sum_{\nu,\rho}\epsilon^{\mu\nu\rho}(d_A
B)^a_{\nu\rho}(x)
}\\
\displaystyle{{{\delta S_{BF,\kappa}}\over{\delta B^a_{\mu}(x)}}}=
\displaystyle{{1\over{4}}\sum_{\nu,\rho}\epsilon^{\mu\nu\rho}
\left(F^a_{\nu\rho}(x)+\kappa^2 f^{abc}B^b_{\nu}(x)B^c_{\rho}(x)\right)
}.
\end{array}
\lab{functder}
\eeq
Assuming that an integration by part can be performed, we have,
for any observable $\Oscr$ classically represented by a $G$-valued function
that transforms under conjugation (or a $\lieg$-valued function
that transforms under the adjoint action), that
\beq
\begin{array}{ll}
\displaystyle{
\la \tr \left[(F +{{\kappa^2}\over2}[B,B])^a_{\mu\nu}(x)
+\kappa (d_AB)^a_{\mu\nu}(x)\right] \Oscr\ra} =
\\
\\
\displaystyle{4\pi i\sum_{\rho}\epsilon_{\mu\nu\rho}\left\{\la
{{\delta \Oscr}
\over{\delta B^a_{\rho}(x)}} +{{\delta \Oscr}
\over{\delta A^a_{\rho}(x)}}
\ra \right\}}.
\end{array}
\lab{intparts}
\eeq
In other words we can replace the terms $d_AB$ and $F+\kappa^2 B\wedge B$
by the functional derivatives with respect to $A$ and, respectively,
$B$.

When we are given a crossing point $x$ in the
(diagram of a) knot $C$ we associate to it four configurations:
$C_{\pm}$, $C_0$ and  $C_{\times}$.

The first two configurations
$C_{\pm}$ correspond to positive and negative crossing points in the diagram,
$C_0$ corresponds to the link obtained by removing the
crossing point in the only orientation-preserving way and, finally,
$C_{\times}$ corresponds to a singular knot where the crossing point
$x$ is a {
\it transversal} double point.

The observable $\tr \Hol(A+\kappa B;C)$ can
be extended to singular knots. In fact all
the v.e.v.'s are regularized by the separation of
the knot $C$ from its
companion $C_f$. Also the same observable can be easily extended to
links as the product of the traces of the above holonomies evaluated
along
the various components of the link.

The framework of quantum field theory suggests that we should study
the effect of
two families of singular deformations applied to the knot $C$ and its
companion $C_f$:
\begin{enumerate}
\item a singular
deformation of the knot $C$ and its companion $C_f,$ concentrated
around a regular point $x$ and characterized by the requirement
that the surface element spanned by this deformation is transversal
to the knot itself. In other words we are ``twisting" the knot, or,
in the terminology used by Kauffman, changing the writhe.
\item a singular deformation of a singular
knot $C_{\times}$ (and of its companion $(C_{\times})_f$) around
a transversal double point  $x$. The effect of this deformation will be to
remove the double point and to create two different non-singular
knots $C_+$ and
$C_-$, depending on the direction of the deformation.
In this case we are assuming that the surface element spanned by
the deformation lies in the same
plane with one of the two tangent vectors to the knot at $x$
and is transversal
to the other one (see \cite{Bru} for a related  approach).
\end{enumerate}

We choose the fundamental representation of $SU(N).$
If we denote by $R^a$
a basis of $Lie(SU(N)),$ normalized so that
$2\tr R^a R^b=-\delta^{ab}$, then we can derive, as a consequence of the fact
that $\{(1/\sqrt{N})\I,(i\sqrt{2})R^a\}$
is an orthonormal basis in the space of complex $n\times n$ matrices,
the well known
Fierz identity, namely:
\beq
2\sum_a R^a \otimes R^a = \sw - (1/N)\I
\lab{fierz1}
\eeq
where $\sw$ denotes the
twist operator $\left(\sw(x\otimes y)=(y\otimes x)\right)$
and $\I$ is the identity.
In components the Fierz identity reads:
\[
2\displaystyle{\sum_a R^a_{ij}R^a_{kl}= \delta_{il}\delta_{kj} -
(1/N) \delta_{ij}
\delta_{kl}}.
\]

We write the Casimir operator in the fundamental representation as:
$\sum_a R^a R^a= c_2\I$ with $c_2=(2N)^{-1}(N^2-1)$.

First we consider a singular infinitesimal
deformation of {\em type 1}.  By integrating by parts we obtain that
\beq
\delta \la \tr \Hol(A+\kappa B)\ra = \mp 4\pi i \kappa c_2
\la \tr \Hol(A+\kappa B)\ra
\lab{def1}
\eeq
where the sign $\mp$ depends on whether, by combining
the orientation of the small
surface bounded by the deformed loop and the orientation of the
the knot,
we obtain the given orientation of the ambient space or
its opposite, respectively.

If we want to consider a finite deformation  of type 1, as opposed to
an infinitesimal one, we can use
the non-abelian Stokes formula introduced in
\cite{Ar}.  The holonomy of a loop bounding a
rectangular surface $\Sigma$ (with initial point $\xnot$), is expressed in
terms
of a path ordered exponential of the surface integral
\beq
\Pscr \exp \int_{\Sigma} dy \Hol^y_{\xnot}(\sigma) F(y)
(\Hol^y_{\xnot})^{-1}(\sigma)
\lab{araf}
\eeq
where $\sigma$ is a path joining $\xnot$ and  $y\in \Sigma$ with a prescribed
pattern.

In the above formula, we now replace
the curvature $F_{\nu,\rho}$, computed w.r.t. the connection $A +\kappa B$,
by the operator
\[
\exp\left\{
\displaystyle{
4\pi i\sum_{\mu}\epsilon_{\mu\nu\rho}
\left(
{
{\delta}
\over{\delta B^a_{\mu}(x)}
} +{
{\delta }
\over{\delta A^a_{\mu}(x)}
}
\right)}
\right\}
\]
and by a succession of integrations by part (see \cite{C}), we can prove that
a positive twisting of the given knot will multiply
the v.e.v. by  a factor $\alpha\equiv\exp( -4\pi i\kappa c_2)$.
In other words, our v.e.v., computed over a
knot-diagram $C^w$ with a given writhe $w$,
transforms, under a change of writhe, as follows:
\beq
\la \tr\Hol(A+\kappa B)(C^{w\pm1})\ra=\alpha^{\pm 1}\la \tr\Hol(A+\kappa B)
(C^{w})\ra.
\lab{regisotop}
\eeq
The formula above follows from the fact that, thanks to
integration by parts, the $n$-th order variation $\delta^n$,
inserts, into the v.e.v., a term
\beq
\sum_{a_1,\dots a_n} R^{a_1}R^{a_2}\cdots R^{a_n}R^{a_n}R^{a_{n-1}}
\cdots R^{a_1}= (c_2)^n \I.
\lab{rn1}
\eeq
Next, we perform an
infinitesimal deformation of {\em type 2} and use
integration by parts again.
Here the $n$-th order variation $\delta^n$
inserts, into the v.e.v., a matrix $S^{(n)}_{i,j,,k,l}\in End({\complessi}^N
\otimes {\complessi}^N),$ given by
\beq
S^{(n)}_{i,j,k,l}=\sum_{a_1,\dots a_n}\left(
R^{a_1}R^{a_2}\cdots R^{a_n}\right)_{i,j}\left(R^{a_1}R^{a_{2}}
\cdots R^{a_n}\right)_{k,l}
\lab{rn2}.
\eeq
By a repeated use of the Fierz identity we obtain that
\[
S^{(n)}=a^{(n)} \sw + b^{(n)}\I
\]
where we have set
\[
\begin{array} {ll}
2a^{(n)}= (N-1)^n2^{-n}N^{-n}- (-1)^n (N+1)^n 2^{-n}N^{-n}\\
2b^{(n)}= (N-1)^n2^{-n}N^{-n}+ (-1)^n (N+1)^n 2^{-n}N^{-n}.\\
\end{array}
\]
The infinitesimal variation is
\beq
\delta\la \tr \Hol(A+\kappa B;C_{\times})\ra = \mp 2\pi i\kappa c_2
\la \tr \Hol(A+\kappa B;C_0)\ra +\displaystyle{
{{2\pi i \kappa}\over N}
\la \tr \Hol(A+\kappa B, C_{\times})\ra}.
\lab{def2}
\eeq
By considering the total variation, $\delta^{tot}$,
(sum of variations of all orders)
we obtain
\beq
\la \tr \Hol(A+\kappa B;C_{\pm})\ra = a^{\pm}
\la \tr \Hol(A+\kappa B;C_0)\ra +\displaystyle{
b^{\pm} \la \tr \Hol(A+\kappa B, C_{\times})\ra},
\lab{def2fin}
\eeq
where we have set
\[
\begin{array}{ll}
a^{\pm} \equiv (1/2)\exp \left\{(\mp 2\pi i\kappa(N-1)/N\}-
\exp \{\pm 2\pi i \kappa
(N+1)/N \right\}
\\
b^{\pm} \equiv (1/2)\exp \left\{(\mp 2\pi i\kappa(N-1)/N\} +
\exp \{\pm 2\pi i \kappa
(N+1)/N \right\}.
\end{array}
\]
By defining $\beta^2=b^-/b^+=\exp(-4 \pi i \kappa/N)$, we finally obtain
a skein  relation:
\beq
\begin{array}{ll}
\beta \la \tr \Hol(A+\kappa B;C_{+})\ra -&\beta^{-1}
\la \tr \Hol(A+\kappa B;C_{-})\ra\\ &=
(\beta A^+ -\beta^{-1} A^-)\la \tr \Hol(A+\kappa B;C_{0})\ra,
\end{array}
\lab{skein1}
\eeq
By combining \rf{regisotop} and \rf{skein1}, we conclude that
$\displaystyle{\la \tr \Hol(A+\kappa B;C)\ra}$ is the HOMFLY polynomial
$P(l,m)$ evaluated at $l= \alpha\beta,\; m=l^{1/N}-l^{-1/N}.$

\section*{VII. Choice of gauge and link-invariants}

In the previous section we showed that, by assuming that integration by parts
is allowed in the functional integral,
$BF$ theory {\it with} a cosmological constant
reproduces knot-invariants
given by the HOMFLY polynomials evaluated at some specific values
of the  variables.

We expect that, by computing the perturbation expansion,
we are able to recover these knot-invariants as power series.
But in order to perform
the perturbation
expansion we need to make one of the (non-equivalent) choices of gauge.

Different gauge-choices produce different expansions
that are recognized to be equal only after some global normalization
factor (that may  be given by a power series) is taken into account.
Moreover, order by order in perturbation theory, one finds, in different
gauges,
 different sets
of Feynman diagrams to be summed over. In conclusion
different gauge-choices may very well lead to the same invariant
in ways that appear completely different.

Let us examine more closely the different choices
of the gauge in perturbative $BF$ theory (with
cosmological constant).

\begin{itemize}
\item Covariant gauge

In this gauge the knot-invariants are expressed as multiple
integrals given by convolutions of kernels of type $l$ and
$v$ (\rf{lv}). In fact, starting from \rf{taylor}, we can write the v.e.v.
of the holonomy operator associated to a knot $C$ as
\[
\la \sum_n \kappa^n \tr \gamma_n(C,\xnot)\ra = \sum_n \kappa^n
V_n(C)
\]
where the coefficients $V_n(C)$ are defined as follows: first we
define
$\displaystyle{\la \tr \gamma_s(C)\ra_j}$ to be the terms
in $\la \tr \gamma_s(C)\ra $
that are
obtained by
inserting exactly $j$ times the vertex term proportional to $B^3$
(this vertex term is multiplied by a factor $\kappa^2$).
Then we define:
\beq
\begin{array}{ll}
V_{2n}(C)\equiv \displaystyle{\sum_{i=0}^n
\la \tr \gamma_{2i}(C,\xnot)\ra_{n-i}
}\\
\\
V_{2n+1}(C)\equiv \displaystyle{\sum_{i=0}^n \la
\tr \gamma_{2i+1}(C,\xnot)\ra_{n-i}
}.
\lab{covexp}
\end{array}
\eeq
Formally, we can prove that all the $V_n(C)$ \cite{C}
are knot-invariants.
In fact let us consider a small deformation of the knot $C$
and, simultaneously, of its companion $C_f$. When we study
the effect
of this deformation on the terms $V_{2n}(C)$ , we collect the different
contributions with the same order in $\kappa$
and obtain:
\beq
\begin{array}{ll}
\delta V_{2n} &
=\sum_{i=0}^{n-1} \left[ \la\tr(\gamma_{2i} F)\ra_{n-i}
+ \la\tr(\gamma_{2i} BB)\ra_{n-i-1}\right] +\\
 &+\sum_{i=0}^n \la\tr(\gamma_{2i-1} d_AB)\ra_{n-i}
+\la\tr(\gamma_{2n}F)\ra_0.
\end{array}
\lab{deltav}
\eeq
Now \rf{deltav} vanishes identically. In fact we have the
following equations:
\[
\begin{array}{ll}
\la\tr\left(\gamma_{2n}F\right)\ra_0=0 \\
\la\tr\left(\gamma_{2i}F\right)\ra_{n-i}+\la\tr
\left(\gamma_{2i}BB\right)\ra=0\\
\la\tr\left(\gamma_{2i-1}d_AB\right)\ra=0,
\end{array}
\]
that are, respectively, a direct consequence of the identities
\bea
\la B(x) B(y) F(z)\ra &=& 0 \lab{1}\\
\la A(x) A(y) F(z)\ra + \la A(x) A(y) B(z) B(z)\ra &=& 0
\lab{2}\\
\la A(x) B(y) (d_A B)(z)\ra &=& 0 \lab{3}
\eea
The proof of the above identities is straightforward:
for instance the r.h.s.\ of \rf{1}
satisfies the following equation (where group factors have been omitted)
\[
\la B(x) B(y) dA(z) + B(x)B(y) A(z) \wedge A(z)\ra = (d_3v)(x,y,z) +
l(x,z)\wedge
l(y,z)=0.
\]
Here the kernels $l$ and $v$ are defined
by \rf{lv} and can be interpreted as
forms on $\left(\reali^3\right)^2$ (of type (1,1))
and, respectively, on $\left(\reali^3\right)^3$ (of type (1,1,1)).
The operator $d_3$ acts on (1,1,1)-forms and produces (1,1,2)-forms
(see \cite{BT,Bar1}). In other words the $d_3$-differential
of the form $v$ compensates the term $l\wedge l$ when the latter form is
restricted to the part of the boundary of the configuration space
$C_4(\reali^3)$ characterized by 2 coincident points.
The proof of \rf{2} and \rf{3} is completely
similar. The variation of $V_{2n+1}(C)$ can be dealt with in a
completely similar way.

Now each term $\la \tr \gamma_{2i}(C)\ra_{n-i}$ gives rise to
multiple integrals involving the kernels $l$ and $v$
\rf{lv}. The kernel $v$ corresponds to
vertex contractions ($B\lolra A\lolra B$
and $A\lolra A\lolra A$), while the kernel $l$ corresponds
to a contraction $A\lolra B$.

In conclusion,  in $\la \tr \gamma_{2i}(C)\ra_{n-i} $
we encounter the following contributions:
an integral with $n$ kernels of type
$v$, an integral with $n-1$ kernels of type $v$ and 2 kernels
of type $l$, $\ldots$,  and finally
an integral with $2i$ kernels of type $l$ and
$n-i$ kernels of type $v$.

A similar computation yields $V_{2n+1}$. In this way we
can represent the coefficients of the HOMFLY polynomial
evaluated as in  section VI, as sums of multiple integrals
given by convolutions of the kernels \rf{lv}.
This representation of the coefficients of the HOMFLY
polynomials is the one considered in the work of Bott and
Taubes \cite{BT}, that is in turn inspired by \cite{GMM1}
and \cite{Bar1}.

\item Holomorphic gauge

The holomorphic and the axial gauge involve a projection onto
a plane. In these gauges
one cannot expect that the coefficients of the perturbative
expansion directly give knot invariants.
The (3-dimensional) diffeomorphism invariance is broken and some correcting
factor must be introduced in order to restore this invariance.
Both the holomorphic and the axial gauge do not have vertex terms.
This implies that, in contrast to the covariant gauge,
the $n$-th term in the perturbative expansion in the variable
$\kappa$ is given by
$\displaystyle{\la \tr \gamma_n(C)\ra}$.

The temporal delta function in \rf{hovev} implies
that we have to consider contractions between a
$B$-field and an $A$-field only when these
fields lie at the same height (in the $t$-variable).
We take these level surfaces, to be transversal to the knot,
for generic, i.e non critical,
times. From \rf{hovev} we see that the forms to be integrated
are given
by
\[
\displaystyle{\bigwedge_i{-2(dz_i-dw_i) \over {(z_i-w_i)}}}
\]
where the pairs of points $(z_i,w_i)$ represent
points of the knots $C$ and on the companion $C_f$,
where contractions occur \cite{FK}.

Perturbatively, the quantum theory is described by a family
of Feynman diagrams $D_P,$
depending on the set $P$ of all possible contractions at a given order in
$\kappa$. To each of these Feynman diagrams, and to each representation
$R$ of the group $G$, we associate a
group factor $W_R(D_P)$. The v.e.v.'s of interest
are given by
\[
\displaystyle{\la\tr \sum_n \kappa^n \gamma_n(C)\ra =W_R(Z_{\epsilon}(C))
}
\]
where $Z_{\epsilon}(C)$ is a diagram-valued function ($\epsilon$ being
the spacing between $C$ and $C_f$). It is possible to
let $C$ approach $C_f$,
by considering (see \cite{AF})
\beq
Z(C) = \lim_{\epsilon\to0^+} e^{-2\kappa(n^+-n^-)\Theta} Z_\epsilon(C),
\eeq
where  $n^\pm$ are the number of critical points (positive and negative)
of the height-function on the knot $C$,
while $\Theta$ denotes the insertion of an isolated chord.

The resulting diagram-valued partition function has been considered by
Kontsevich \cite{Konts, Bar2}:
\beq
Z(C) = \sum_{m=0}^\infty (-2\kappa)^m
\int_{t_{\rm min}<t_1<\cdots<t_m<t_{\rm max}}
\sum_{P=\{(z_i,z_i')\}} (-1)^{\# P_\downarrow} D_P
\bigwedge_{i=1}^m \frac{dz_i-dw_i}{z_i-w_i},
\lab{Kinvariant}
\eeq
where
$t_{\rm min}$ and $t_{\rm max}$ denote the lowest and highest height
of $C$, respectively,
$(z_i,t_i)$ and  $(w_i,t_i)$ denote distinct points on $C$
and $\# P_\downarrow$ denotes the number of points $(z_i,t_i)$,$(w_i,t_i)$
where the height is a decreasing function.

In \cite{Bar2},
it is shown that such integrals are well-defined
knot invariants, provided that we use the normalization:
\beq
\hat Z(C) = \frac{Z(C)}{Z(\infty)^{\frac c2-1}},
\eeq
where $c$ is the number of critical points
and $\infty$ denotes the particular unknot with one crossing point
whose diagram looks like the symbol $\infty$.

\item Axial gauge

The v.e.v.'s in this gauge (see eq. \rf{axvev}) indicate that the interactions
to be considered are localized only at the crossing points
of the projected knot $C$ ($B$-field line)
with its framing $C_f$ ($A$-field line).
These crossing points occur
either when $C_f$ is ``twisted" around $C$ or when we have an actual
crossing point of the projected knot $C$.

There is an obvious invariance of v.e.v.'s under
orientation-preserving transformations of the plane (or, more generally,
of the surface $\Sigma$) onto which
the knot is projected. Hence  these interactions
only depend on the type of the crossing (positive or negative)
of the projected
knot $C$ with its companion $C_f$.

We have first to take into consideration a change of the framing (i.e.
a twisting of $C_f$). Once we have done this, we can then choose a specific
framing (the ``blackboard framing") where $C_f$ is always at the
right of $C$ (with respect to the given orientation of $C$)
and hence intersects
$C$ only near an actual vertex of the knot $C$. In this way, only actual
vertices of $C$ contribute to the interaction.
After having chosen the framing as above, we explicitly compute
$\displaystyle{\la \tr \sum_n \kappa^n \gamma_n(C)\ra }$.

At the $n$-th order of perturbation we have $n$ interactions localized
at the vertices. These interactions will be
represented by
(traces of) group factors. For the fundamental representation
of $SU(N)$, they are given by functions of $N$.

Moreover, among the $n$ interactions that we are considering,
$n_1$ of them can be localized at one vertex, $n_2$
at another vertex and $n_j$ at a $j$-th vertex. The requirement here
is that
$\sum_j n_j=n$.

In other words, we can write the $n$-th term in the perturbation expansion
as a sum
\beq
\begin{array}{ll}
\sum_{i_1,i_2,\cdots,i_n}\epsilon_{i_1}\epsilon_{i_2}\cdots
\epsilon_{i_n}
D_{n,n}^{i_1,i_2,\cdots,i_n}(C)+ &
\sum_{i_1,i_2,\cdots,i_{n-1}}\epsilon_{i_1}\epsilon_{i_2}\cdots
\epsilon_{i_{n-1}}
D_{n,n-1}^{i_1,i_2,\cdots,i_{n-1}}(C)\\   &+\cdots +
\sum_i \epsilon_i D_{n,1}^i(C),
\end{array}
\lab{homflyax}
\eeq
where the indices $i_j$ label the vertices of the knot $C$ and $D_{n,i}$ are
(traces of) group factors corresponding to $n$ interactions concentrated
at $i$ different vertices.

We cannot hope to obtain  the HOMFLY polynomial directly
from \rf{homflyax},
since we have had to make a particular choice of the framing (blackboard
framing).

But a good choice of the normalization factor
for \rf{homflyax} will show that the
perturbation theory in the axial gauge (with the fundamental representation
of $SU(N)$)
provides the expression of
the coefficients of the HOMFLY polynomials in terms of tensors
over the vertices of the knot (see
the discussion in the appendix).

The similarity between
\rf{homflytens} and \rf{homflyax} is striking.

\end{itemize}

\section*{IX. The $BF$ theory without cosmological constant and the
Alexander--Conway polynomial}

In this section we consider the $BF$ theory {\it without} cosmological
constant. The observable associated to a knot $C$ is then given
by $\tr \Hscr(C;\lambda)$, where $\lambda$ is an expansion parameter and
$\Hscr$ is defined by \rf{chenab} and \rf{taylorab}.

Another observable that could be considered in $BF$ theory
without cosmological constant is
$\tr \exp\left(\lambda \Gamma_1(C;\xnot)\right)$,
the difference between the v.e.v.'s computed
for the two observables being given by powers of $lk(C_f,C)$.
In \cite{CCM} the latter definition of the observable was assumed;
but the choice $\tr \Hscr(C;\lambda)$ is  a more
natural one if one wants to deal with arbitrary values of
$lk(C_f,C)$.

For simplicity we assume that $lk(C_f,C)=0$ (standard framing);
this makes the distinction between the two choices of observables
irrelevant and allows us to use the results of \cite{CCM}.

The perturbation expansion in the covariant gauge reads
\[
\la \sum_n \lambda^n \tr \Gamma_n(C,\xnot)\ra = \sum_n \lambda^n W_n(C)
\]
where the coefficients $W_n(C)$ are defined as
\beq
W_n(C) = \la \tr \Gamma_n(C,\xnot) \ra.
\lab{W}
\eeq
In contrast to \rf{covexp},  we do not have to take into account
the effect of vertex terms proportional to $B^3$
in eq.\ \rf{W}. Hence the structure
of $BF$ theory {\it without} cosmological constant is considerably
simpler than the one of the $BF$ theory {\it with} a cosmological constant.

We now formally prove that the terms $W_n(C)$ are knot-invariants.

In fact, the effect of a small deformation of the knot $C$ is
given by
\beq
\delta W_{2n}
=\sum_{i=0}^{n} \la\tr(\Gamma_{2i} F)\ra
 +\sum_{i=0}^n \la\tr(\Gamma_{2i-1} d_AB)\ra=0.
\lab{deltaw}
\eeq
In order to prove \rf{deltaw},
we have used again equations \rf{1} and \rf{3}, while equation
\rf{2} did not play any r\^ole, since,
here, there are no vertex terms proportional to $B^3$,
and hence no terms like $\la A^3\ra$.

When we consider a knot $C$ ({\it not} a link) and when
the standard framing for $C_f$ is selected, then it has been shown in
\cite{CCM} that
\beq
W_{2n+1}=0
\lab{disparizero}
\eeq
holds for any $n.$

As far as integration by parts for $BF$ theories without cosmological
constant is concerned, formulas \rf{functder} still hold, provided
that $\kappa$ is set equal to zero. Formula \rf{intparts}
becomes
\beq
\begin{array}{ll}
\displaystyle{
\la \tr \left[F^a_{\mu\nu}(x)\Oscr\right]\ra}=
\displaystyle{4\pi i \sum_{\rho} \epsilon_{\mu\nu\rho}
\la
{{\delta \Oscr}
\over{\delta B^a_{\rho}(x)}}
\ra
}
\\
\\
\displaystyle{
\la \tr \left[(d_AB)^a_{\mu\nu}(x) \Oscr\right]\ra} =
\displaystyle{
4\pi i \sum_{\rho} \epsilon_{\mu\nu\rho}
\la
{{\delta \Oscr}
\over{\delta A^a_{\rho}(x)}}
\ra }.
\end{array}
\lab{intpartsalex}
\eeq
We consider, once again, a deformation
of the knot $C$ and, simultaneously, of its framing $C_f$, while keeping
$lk(C_f,C)=0$.
We redo the computations of section VI by using integration
by parts and the abelian Stokes formula.

We do not need to take into consideration deformations of type 1
(see  section VI), since the imposition of
the requirement $lk(C,C_f)=0$ will offset the effect of such
deformations.

When we consider deformations of type 2 (see section VI) then
the form of our observables shows that the $B$-field  is not path-ordered
any more. Hence, instead of \rf{rn2}, the $n$-th order variation
inserts, at the selected crossing point, the matrix
 $U^{(n)}_{i,j,k,l}\in End({\complessi}^N
\otimes {\complessi}^N),$ given by
\beq
U^{(n)}_{i,j,k,l}={1\over{n!}}
\sum_{a_1,\dots a_n;\sigma}\left(
R^{a_1}R^{a_2}\cdots R^{a_n}\right)_{i,j}\left(R^{a_{\sigma(1)}}R^{a_
{\sigma(2)}}
\cdots R^{a_{\sigma(n)}}\right)_{k,l}
\lab{rn2ale},
\eeq
where $\sigma$ denotes a permutation of $\{1,2,\cdots,n\}$ and the sum
is extended over all permutations and over the indices $\{a_j\}$.
It is possible to show that \rf{rn2} is still a matrix of the form
$\alpha_n \sw +\beta_n \I$ \cite{mr}. This implies that
we still obtain a relation like \rf{def2fin}:
\[
\la\tr \Hscr(C_{\pm};\lambda)\ra=\alpha^{\pm} \la\tr \Hscr(C_0;\lambda)\ra
+\beta^{\pm} \la\tr\Hscr(C_{\times};\lambda)\ra
\]
that, in turn, implies a skein relation that we write as:
\beq
q(\lambda)\sum\lambda^n W_n(C_+) - q^{-1}(\lambda)\sum\lambda^nW_n(C_-) =
z(\lambda)
\sum\lambda^nW_n(C_0).
\label{skeinalex}
\eeq
In conclusion, the v.e.v. associated to a knot $C$ (or a link)
in the $BF$ theory
without cosmological constant assigns a skein polynomial
\[
P(q(\lambda),z(\lambda))(C)
\]
to $C$.
In order to identify this polynomial $P(q(\lambda),z(\lambda))$
with the Alexander--Conway polynomial, we need one further observation:
the transformation
\[ \lambda\lora -\lambda\]
can be absorbed, in field theory, by the transformation $ B\lora -B$
that, in turn, is equivalent to a change in the sign of the $BF$ action,
or to a change in the {\it orientation} of the manifold $M$ ($=\reali^3$).

So we have that
\beq
P(q(-\lambda),z(-\lambda))(C)=P(q(\lambda),z(\lambda))(C^{!})=
P([q(\lambda)]^{-1},-z(\lambda))(C)
\label{mirror}
\eeq
where $C^{!}$ denotes the mirror image of $C$.
The first identity in \rf{mirror} is a
consequence of the property (CT-symmetry) of field theory mentioned
above,while the second identity is a consequence of the skein relation.

We now take a {\it knot} $C$, for which \rf{disparizero} implies that
$P(q(\lambda),z(\lambda))(C)=P(q(\lambda),z(\lambda))(C^{!})$.

Hence, for a knot $C$,
we have $q(\lambda)=1$ and, when we choose the
normalization $P(1,z(\lambda))(\bigcirc)=1$ for the unknot, then
$P(1,z(\lambda))(C)$
must necessarily be the {\it Alexander--Conway polynomial}.

$BF$ theory without cosmological constant yields the Alexander--Conway
polynomial
also in the case of links: but for the discussion
of this case we refer to \cite{CCM}.

\section*{X. Observables for the four-dimensional $BF$ theory}

Our purpose, in this section, is only to sketch a few preliminary ideas
on how to deal with 4-dimensional situations, leaving
further developments to be carried out elsewhere.

As has been mentioned in Section III, the observables associated to
a 4-dimensional $BF$ theory must be associated to 2-dimensional
surfaces $\Sigma$ imbedded (or immersed) in the 4-manifold $M$.
As in \cite{CM}, we can associate to
$\Sigma$ (with a selected
point $\xnot$) the quantity:
\beq
\displaystyle{\tr \int_M \Hol(A)_{\xnot}^y B(y)
\Hol(A)^{\xnot}_y}.
\lab{bobs}
\eeq
In this formula the holonomies are meant to be computed along
loops with base-point $\xnot,$ passing through the
point $y\in \Sigma$.

What we would like to do is
to associate to each point of the surface $\Sigma$ a loop with base-point
$\xnot$. This can be a difficult task, if we want to preserve
smoothness\rlap.\footnote{We
acknowledge a useful discussion with John Baez, in this respect.}

A simpler situation is encountered if we are given an oriented torus
$\toro=S^1\times S^1$. The torus $\toro$ is imbedded
in $M$ (or, more generally,
generically immersed  i.e. with only transversal double points).

We still denote by $\Sigma$ the image of such
imbedding (or immersion). Here we can define, as in \cite{Ar}, a special
path joining $\xnot$ to the generic point $y\in \toro$. If the
coordinates of the points $\xnot, y$ are $(s_0,t_0),(s,t)\in \toro,$
then we define a path $\sigma_y$
by combining a (positively oriented)
meridian arc  joining $(s_0,t_0)$ to $(s,t_0)$ with
a (positively oriented)
longitudinal arc joining $(s,t_0)$ to $(s,t)$.

As in section III, we consider the $\lieg$-valued 2-form of the
adjoint type:
\beq
\hat B(y) \equiv \Hol(A,\sigma_y)_{\xnot}^y B(y) \left[
\Hol(A,\sigma_y)_{\xnot}^y\right]^{-1}
\lab{hatb4}
\eeq
and consider
\[
\displaystyle{\int_{\Sigma} \hat B(y)}
\]
Moreover, we can ``complete" the above holonomies and obtain:
\beq
O(\Sigma)\equiv \displaystyle{\int_{\Sigma} \hat
B(y)\Hol(A)_{s_0,t_0}^{s,t_0}\Hol^{(l)}_{s,t_0}
\Hol_{s,t_0}^{s_0,t_0}
}
\lab{hol4}
\eeq
where by $\Hol^{(l)}_{s,t_0}$ we mean the holonomy of the longitudinal
circle with base point $(s,t_0)$.

At this point we can exhibit the observable for the 4-dimensional
$BF$ theory, namely
\beq
\Oscr(\Sigma, k)\equiv \tr \exp\left(k\ O(\Sigma)\right).
\label{O4}
\eeq

This is an observable for the 4-dimensional $BF$ theory without
cosmological constant.
We can now compute
the relevant v.e.v.'s by perturbation expansion in the
coupling constant $k$.
The gauge-choices that we have here
at our disposal are
either the covariant gauge or the real axial gauge.

$BF$ theory in 4 dimensions should provide the right framework
for invariants of 2-knots (embedded surfaces)
or of singular 2-knots (generally immersed surfaces).

Preliminary computations (see \cite{CM}) suggest that the expression
of these invariants in the covariant gauge, should be given
in terms of iterated integrals of kernels that are the higher dimensional
generalization of \rf{lv}.

Preliminary computations in another direction, show that the observable
\rf{O4} could play a r\^ole in the recovery of some essential information
concerning the differentiable structure of the four manifold $M$
(Donaldson polynomial) in the
framework of a pure Yang--Mills theory \cite{CCGM}.

Finally let us point out that the observable \rf{O4} can be a relevant
object in the approach to quantum gravity based on loop-variables
(see \cite{ARS} and reference therein).
In this framework,  imbedded (or generically
immersed) surfaces (or tori) represent the time-evolution of the
loop variables of \cite{ARS}.
An elementary consideration shows moreover that, once we are
given a background metric $g$ (and so a corresponding *-operator)
the $B$-field can provide a fluctuation of the background metric.

In fact one can construct a symmetric tensor
\[
h_{\mu,\nu} \equiv f_{a,b,c}
B^a_{\mu,\rho} {^*}(B^b)^{\rho,\sigma}
B^c_{\sigma,\nu}
\]
where $f^{a,b,c}$ denote the structure constant and a sum over repeated
indices is understood (see \cite{Cap}).

We will discuss more about the 4-dimensional $BF$ theory in  future
work.

\section*{Acknowledgements}
We thank John Baez and Maurizio Rinaldi for very useful
discussions. Two of us (A.S.C. \& M.M.) thank
E.T.H. Z\"urich for hospitality.

\section*{Appendix: On the coefficients of the skein-polynomials}

In this appendix we want to review the
representation of the coefficients of the skein-polynomials
in terms of suitably defined ``tensors" with coefficients in $\interi$.

The motivation for this appendix is that this representation is the one
obtained
from the quantization of the $BF$ theories in the (real)
axial gauge, with the fundamental
representation  of the group $SU(N)$.

First we consider a  link diagram $L$ with $|L|$ oriented components,
and we denote by $V(L)$
the set of vertices of $L$ (crossing points).
We order the set $V(L)$ by ordering
arbitrarily the components of $L$ and by
choosing arbitrarily a starting point in each component
of $L$. We denote the sign (writhe)
of the $i$-th vertex by $\epsilon_i=\pm 1 $.
A $k$-tensor $T\equiv T^{i_1,i_2,\cdots,i_k}$
is defined as a map
\[
\underbrace{
V(L)\times\cdots\times V(L)}_{\mbox{$k$ times}}\lora \interi.
\]
Once we are given a $k$-tensor $T,$ we can ``saturate" it with the writhe
of the vertices in $V(L)$
\beq
\sum_{V(L)\times \cdots \times V(L)}\epsilon_{i_1}\epsilon_{i_2}\cdots
\epsilon_{i_k}
T^{i_1,i_2,\cdots,i_k}.
\lab{tens}
\eeq
Here we want to show that the coefficients of the skein polynomials
are given by sums of expressions \rf{tens}.

We denote by $S_j$ the operation of switching the vertex $v_j\in V(L)$ (i.e.
\ changing the writhe) and by $E_j$ the operation of
eliminating the vertex $v_j$ in the only orientation-preserving way.
Let $\sigma$ be any sequence of the above operations. Given a
$k$-tensor $T$ on $\sigma L$ we can pull it back to a $k$-tensor on $L$
by defining:
\[
\begin{array}{ll}
[\sigma^*(T)]^{i_1,i_2,\cdots,i_k}= \cases
{0 & if one of the vertices $v_{i_r}$\\
& has been eliminated,
\\ \rho^{\sigma}(v_{i_1},v_{i_2}, \cdots,
v_{i_k}) T^{\sigma(i_1),\sigma(i_2),\cdots,\sigma(i_k)}
& if none of the vertices $v_{i_r}$ \\
& has been eliminated,\\}
\end{array}
\]
where
$\rho^{\sigma}(v_{i_1},v_{i_2},\cdots,v_{i_k})$ is defined
as $1$ if an even number
of the vertices $v_{i_1}, \cdots, v_{i_k}$ has been switched by $\sigma$
and is defined as $-1$ otherwise.
The pulled-back $k$-tensor satisfies the relation:
\beq
\sum_{i_1,i_2,\cdots,i_k}\epsilon_{i_1}\epsilon_{i_2}\cdots
\epsilon_{i_k}
[\sigma^*(T)]^{i_1,i_2,\cdots,i_k}(L)
=\sum_{j_1,j_2,\cdots,j_k}\epsilon_{j_1}
\epsilon_{j_2}\cdots
\epsilon_{j_k}
T_n^{j_1,j_2,\cdots,j_n} (\sigma L),
\lab{pulled}
\eeq
where the first sum is extended over all the k-uples
of vertices of $L$, while the second sum is extended over all the k-uples
of vertices of $\sigma(L)$.

We now consider the Alexander--Conway polynomial
$\Delta(L)(z)\equiv \sum_n a_n(L)z^n$, with the standard normalization
conditions.

Let $\{v_{j_1},v_{j_2},\cdots,v_{j_s}\}$ be any sequence of
vertices
of $L$ with the property that when we
switch all these vertices then the link diagram
$L$ is transformed into the diagram of the
unlink $U_{|L|}$ (with $|L|$ components). We have (\cite{BM, Kau}):
\beq
\begin{array}{ll}
a_n(L)-a_n(S_{j_1}L)=\epsilon_{j_1}a_{n-1}(E_{j_1}L)\\
a_n(S_{j_1}L)-a_n(S_{j_2}S_{j_1}L)=\epsilon_{j_2}a_{n-1}(E_{j_2}S_{j_1}L)\\
\cdots\cdots\\
a_n(S_{j_{s-1}}\cdots S_{j_1}L)-a_n(S_{j_s}S_{j_{s-1}}\cdots S_{j_1}L)=
\epsilon_{j_s}a_{n-1}
(E_{j_s} S_{j_{s-1}}\cdots S_{j_1}L).
\end{array}
\lab{coeff}
\eeq
By assumption we have that $S_{j_s}S_{j_{s-1}}\cdots S_{j_1}L=U_{|L|}$ and
hence
\[
\begin{array}{ll}
a_n(S_{j_s}S_{j_{s-1}}\cdots S_{j_1}L)=\cases
{1 &
if  $|L|=1$ and $n=0$ \\
0 & otherwise.\\}
\end{array}
\]
When $n>1$ we obtain:
\beq
a_n(L)= \sum_{l=1}^{l=s}
\epsilon_{j_l}a_{n-1}(E_{j_l}S_{j_{l-1}}\cdots S_{j_1}
L).
\lab{an}
\eeq
We are now ready to prove, by induction, that
for any link $L$, the $n$-th coefficient of the Alexander polynomial
is given by an expression like:
\beq
a_n(L)=\sum_{i_1,i_2,\cdots,i_n}\epsilon_{i_1}\epsilon_{i_2}\cdots
\epsilon_{i_n}
A_n^{i_1,i_2,\cdots,i_n}(L),
\lab{alex}
\eeq
 where $A_n(L)$
is a suitable $n$-tensor with integer entries
and the sum is extended over all the $n$-uples of vertices
in $L$.

For $n=1$, we have
$a_1(L)\equiv \sum_j\epsilon_jA^j(L),$ where the 1-tensor $A^j$ is defined as
\beq
\begin{array}{ll}
A^j(L)=\cases{1 & if $|L|=2$ and the first component passes {\it over}\\
& the second
one at the $j$-th vertex\\
0 & otherwise\\}.
\end{array}
\lab{lk}
\eeq
Equation \rf{an} can be rewritten as a sum extended over {\it all} the
vertices of $L$:
\beq
a_n(L)= \sum_i \epsilon_i{\tilde a}^i_{n-1},
\lab{an2}
\eeq
where ${\tilde a}^i_{n-1}$ is defined as
either 0 (when the $i$-th vertex is not one of
vertices $v_{j_l}$
that we need to switch in order to transform $L$ into the unlink)
or is given by \rf{an}.
We now assume that
$a_{n-1}(E_{j_l}S_{j_{l-1}}\cdots S_{j_1})$ can be expressed in terms
of an $(n-1)$-tensor of the link $(E_{j_l}S_{j_{l-1}}\cdots S_{j_1})$.
This implies that it can  also be expressed
in terms of an $(n-1)$-tensor of the link $L$, and  so eq.~\rf{an2}
directly
gives  \rf{alex}.

We finally consider the 1-variable HOMFLY Polynomial $P(\exp(hN),2\sin(h))
$.
We represent this polynomial as a power series in the variable $h$,
namely as $\sum_{n=0}^{\infty} a_n h^n$, where it is understood that
the coefficients $a_n$ depend on the link $L$ and on the integer $N$.
We choose the following normalization condition for the unlink with $k$
components
\beq
P(U_k)= \displaystyle{\left({{\exp(hN)-\exp(-hN)}\over {\exp(h)-\exp(-h)}}
\right)^{k}}.
\lab{norm}
\eeq
The skein relation and eq.~\rf{norm}
imply that for, any link $L,$ we have $a_0(L)=N^{|L|}$.
We also have
$a_1(U_k)=0$, and, for any vertex $v_j$,
\[
a_1(L)-a_1(S_jL)=\epsilon_j 2(N^{|E_jL|}-N^{|L|+1}).
\]
This implies that if
$L$ is a link with zero linking numbers between its components,
then $a_1(L)=0$. More generally, for a link $L$ with $k>1$ components,
we have:
\[
a_1(L)= \sum_s \epsilon_{j_s} 2(N^{k-1}-N^{k+1})
\]
where $\{v_{j_s}\}$
is a set of vertices, where different components of $L$ meet
and whose switching separates the components.

For the generic coefficient $a_n$ we
now have a set of equations that  is more complicated
than \rf{coeff}
\beq
\begin{array}{ll}
\sum_{k=0}^{n}(n-k)!&\left[N^{n-k}a_k(L)-(-N)^{n-k}a_k(S_{j_1}L)\right]=
\epsilon_{j_1}\sum_{k=0}^{n-1}(n-k)![1-(-1)^{n-k}]a_{k}(E_{j_1}L)\\
\cdots\cdots\\
\sum_{k=0}^{n}(n-k)!
&\left[(N^{n-k}a_k(S_{j_{s-1}}\cdots
S_{j_1}L)-(-N)^{n-k}a_k(S_{j_s}S_{j_{s-1}}\cdots S_{j_1}L)\right]=
\\
&\epsilon_{j_s}\sum_{k=0}^{n-1}(n-k)![1-(-1)^{n-k}]
a_{k}(E_{j_s} S_{j_{s-1}}\cdots S_{j_1}L).
\end{array}
\lab{coeff1}
\eeq
Here $\{v_{j_s}\}$
is a set of vertices of $L$
whose switching transforms $L$ into the unlink $U_{|L|}.$

By an argument similar to the one considered before, instead
of \rf{alex} we now have equations like:
\beq
\begin{array}{ll}
a_n(L)=\sum_{i_1,i_2,\cdots,i_n}\epsilon_{i_1}\epsilon_{i_2}\cdots
\epsilon_{i_n}
A_{n,n}^{i_1,i_2,\cdots,i_n}(L) +&\sum_{
i_1,i_2,\cdots,i_{n-1}}\epsilon_{i_1}\epsilon_{i_2}\cdots
\epsilon_{i_{n-1}}
A_{n,n-1}^{i_1,i_2,\cdots,i_{n-1}}(L)\\
 &+\cdots +
\sum_i \epsilon_i A_{n,1}^i(L),
\end{array}
\lab{homflytens}
\eeq
where to each index $n$ we associate tensors $A_{n,i}$ of order $i$
for $i=1,\cdots,n$. In \rf{homflytens} the sums  are extended
over the set $V(L)$.


\begin{thebibliography}{99}

\bibitem{AB} M.F.Atiyah, R.Bott {\em The Yang-Mills Equations over Riemann
Surfaces} Phil. Trans. R. Soc. Lond. {\bf A} (1982) 523-615

\bibitem{AF} D.Altschuler, L.Freidel {\em On Universal Vassiliev Invariants}
hep-th/9403053


\bibitem{Ar} I.Ya.Aref'eva {\em Non-abelian Stokes formula}
Teor. Math. Fiz. {\bf 43} 111-116

\bibitem{ARS} A.Ashtekar, C.Rovelli, L.Smolin {\em Weaving a
Classical Geometry with Quantum Threads} Phys. Rev. Lett. {\bf 69}
(1992) 237-240

\bibitem{BM} R.Ball, M.L.Mehta {\em Sequence of Invariants for knots
and Links} J. Physique {\bf 42} (1981) 1103-1199

\bibitem{Bar1} D.Bar-Natan {\em Perturbative aspects of the
Chern--Simons field theory} PhD Thesis, Princeton University (1991)

\bibitem{Bar2} D.Bar-Natan {\em On Vassiliev's Knot Invariants},
Topology (to appear)

\bibitem{Bla} D.Birmingham, M.Blau, M.Rakowski, G.Thompson
{\em Topological Field theories}, Phys. Rep. {\bf 209} (1991) p.129-340

\bibitem{Bru} B. Br\"ugmann, {\em On a geometric derivation of Witten's
identity for Chern--Simons theory}, preprint MPI-PH-93-107,
hep-th/9401055.

\bibitem{BT} R.Bott, C.Taubes {\em On the self-linking of knots}
J. Math. Phys. {\bf 35} (1994) 5247-5287

\bibitem{C} A.S.Cattaneo {\em Teorie topologiche di tipo $BF$ ed
invarianti dei nodi} PhD Thesis, University of Milan (February 1995)

\bibitem{Cap} R.Capovilla, T.Jacobson, J.Dell {\em General
relativity without the metric} Phys. Rev. Lett. {\bf 63} (1989)
2325-2328

\bibitem{CCM} A.S.Cattaneo, P.Cotta-Ramusino, M.Martellini,
{\em Three-dimensional $BF$ Theories and the Alexander--Conway Invariant
of Knots} Nucl. Phys. {\bf B 436}, 355-384

\bibitem{CCGM} A.S. Cattaneo, P.Cotta-Ramusino, A. Gamba, M.Martellini
{\em The Donaldson-Witten Invariants and Pure QCD with Order and Disorder
't Hooft-like Operators} Milan University preprint (1995)
IFUM 493/FT, hep-th/9502110

\bibitem{Chen} K.T.Chen {\em Iterated Path Integrals} Bull. Am. Math.
Soc. {\bf 83} (1977) 831-879

\bibitem{CM} P.Cotta-Ramusino, M.Martellini, {\em BF-theories and 2-knots}
in {\em Knots and Quantum Gravity}, J.Baez (ed.), Oxford University Press,
Oxford, New York (1994) 169-189

\bibitem{CGMM} P.Cotta-Ramusino, E.Guadagnini, M.Martellini,
M.Mintchev, {\em Quantum field theory and link invariants}, Nucl. Phys.
{\bf B330} p. 557-574 (1990)

\bibitem{Weber} P.De La Harpe, M.Kervaire, C.Weber {\em On the Jones
Polynomials}, L'Ens. Math. {\bf 32} (1986) 271-335

\bibitem{FK} J.Fr\"ohlich, C.King {\em The Chern--Simons Theory and
Knot-Polynomials} Comm. Math. {\bf 126} (1989) 167-199

\bibitem{GMM1} E.Guadagnini, M.Martellini, M.Mintchev, {\em
Chern--Simons model and new relations between the
HOMFLY coefficients}, Phys. Lett. {\bf B228},
p. 489-494 (1989)

\bibitem{HOMFLY} P. Freyd, D. Yetter, J. Hoste, W.B.R. Lickorish, K.C. Millet,
A. Ocneanu {\em A new polynomial invariants of Knots and Links} Bull.
Am. Math. Soc.
{\bf 12} (1985) 239-246

\bibitem{Jones} V.F.R. Jones {\em A polynomial invariant for links
via Von Neumann algebras} Bull. Am. Math. Soc. {\bf 12} (1985) 103-112

\bibitem{Kau} L.H.Kauffman {\em On Knots} Princeton University Press
(1987), Princeton N.J.

\bibitem{Kauff} L.H. Kauffman {\em The Conway Polynomial}, Topology
{\bf 20}  (1981) 101-108

\bibitem{Konts} M.Kontsevich {\em Vassil'ev Knot Invariants}
Adv. Sov. Math. (1992) {\bf 16} 137-150

\bibitem{Lic} W.B.R. Lickorish, K.C. Millet {\em A polynomial invariant of
Oriented links}, Topology {\bf 26}, p.107-141 (1987)

\bibitem{mr} M.Rinaldi, private communication

\bibitem{RT} N.Reshetikhin, V.G.Turaev  {\em Invariants of 3-Manifolds via Link
Polynomials
and Quantum Groups}
Invent. Math. {\bf 103} (1991) 547-598

\bibitem{Schw} A.S. Schwarz {\em The Partition Function of
a degenerate quadratic functional and the Ray-Singer invariants}
Lett. Math. Phys. {\bf 2} (1978) 247-252

\bibitem{TV} V.G.Turaev, O.Yu.Viro {\em State Sum Invariants of 3-Manifolds
and Quantum 6j-symbols} Topology {\bf 31} (1992) 865-902

\bibitem{Wit1} E.Witten  {\em Topological Quantum
Field Theory} Comm. Math. Phys. {\bf 117} (1988) 353-386

\bibitem{Wit2} E.Witten {\em Quantum Field Theory and the Jones
polynomials} Comm. Math. Phys. {\bf 121} (1989) 351-399

\end{thebibliography}
\end{document}